\documentclass{ws-procs975x65}

\begin{document}

\title{\Large Higher Dimensional Rotating Charged Black Holes}

\author{ALIKRAM  N. ALIEV\footnote{aliev@gursey.gov.tr}}

\address{Feza G\"ursey Institute, P.K. 6  \c Cengelk\" oy, 34684 Istanbul, Turkey}

\begin{abstract}
We discuss a new analytic solution to the Einstein-Maxwell field
equations that describes electrically charged black holes with a
slow rotation and  with a single angular momentum in all higher
dimensions. We also compute the gyromagnetic ratio of these black
holes.
\end{abstract}

\bodymatter

\section{Introduction}
Black holes remain to be one of the most intriguing and puzzling
object of study in higher dimensional spacetimes. It is widely
believed that the remarkable features of black holes in four
dimensions, such as the equilibrium and uniqueness properties as
well as quantum properties of evaporation of mini-black holes
might have played a profound role in understanding the nature of
fundamental theories in higher dimensions. Therefore, of
particular interest is the study of black hole solutions in
higher-dimensional gravity theories as well as in string/M-theory.

The first higher dimensional solutions for static black holes with
the spherical topology of the horizon have been discussed by
Tangherlini \cite{tang}. These solutions generalize the
spherically symmetric Schwarzschild and Reissner-Nordstrom
solutions of four-dimensional general relativity. For the static
black holes the uniqueness and the stability properties still
survive \cite{gibbons1} in higher dimensions, however the
situation is different for rotating black holes. The exact
solution for the rotating black holes was found by Myers and Perry
\cite{mp}. The solution is not unique, unlike its four dimensional
counterpart, the Kerr solution. There exists a rotating black ring
solution \cite{er} in five dimensions with the horizon topology of
$  S^2 \times S^1 $ which may have the same mass and angular
momentum as the Myers-Perry solution. However,  the counterparts
of the Myers-Perry black holes in higher dimensional
Einstein-Maxwell theory have not been found yet. A numerical
treatment of the problem for some special cases in five dimensions
was given in Ref. \cite{kunz1}. Here, as a first step towards the
desired exact metric,  we shall discuss the intermediate case of
{\it slow rotation} and give  a new analytical solution that
describes electrically charged black holes with a single angular
momentum in all higher dimensions.

\section{Metric ansatz}
The strategy of obtaining the familiar Kerr-Newman solution in
general relativity  is based on either using the metric ansatz in
the Kerr-Schild form or applying the method of complex coordinate
transformation to a non-rotating charged black hole. When
employing in $ N+1 $ dimensional spacetime  both approaches lead
us to the following metric ansatz
\begin{eqnarray}
ds^2 & = &- \left(1-\frac{m}{r^{N-4} \,\Sigma}+
\frac{q^2}{r^{2(N-3)} \,\Sigma} \right) dt^2 - \frac{2 a\left(m
r^{N-2}-q^2\right)
\sin^2 \theta}{r^{2(N-3)}\,\Sigma}\,dt \,d\phi \nonumber \\[2mm] &&
 + \left(r^2+a^2+
\frac{a^2 \left(m r^{N-2}-q^2\right) \sin^2\theta}
{r^{2(N-3)}\,\Sigma}\right)\sin^2{\theta}\,d\phi^2 + {{r^{\,N-2}\,
\Sigma }\over \Delta}\,dr^2 \nonumber \\[2mm] &&
+ \Sigma \, d\theta^2 + r^2 \cos^2{\theta}  d\Omega_{N-3}^2\,\,,
\label{ansatz}
\end{eqnarray}
where the parameters $ m $ , $a $  and $ q $ are related to the
mass, angular momentum and electric charge of the black hole. The
metric function
\begin{equation}
\Delta= r^{N-2}(r^2 + a^2) -m \, r^2 + q^2\, r^{4-N}\,\,
\end{equation}
and $ d\Omega_{N-3}^2 $ is the metric on a unit $(N-3) $-sphere.

It is straightforward  to verify that the source-free Maxwell
equations in the background of the metric (\ref{ansatz}) admit the
potential one-form field \cite{aliev1}
\begin{equation}
A= -\frac{Q}{(N-2)\, r^{N-4}\,\Sigma}\left(dt- a
\sin^2\theta\,d\phi \right)\,\,. \label{potform2}
\end{equation}
where $ Q $ is the electric charge of the black hole. For $ N=3 $,
inspecting the  simultaneous system of the Einstein-Maxwell
equations with this potential form and with the metric
(\ref{ansatz}) shows that it is satisfied for $ q^2=G Q^2 $. This
is the case of a Kerr-Newman black hole in four dimensions.
However, for $ N \geq 4 $, the consistent solution to the system
of equations is obtained only when we restrict ourselves to a slow
rotation.
\section{Results}
The metric for slowly rotating and charged black holes with a
single angular momentum in all higher dimensions has the form
\begin{eqnarray}
ds^2 & = &- \left(1-\frac{m}{r^{N-2}}+ \frac{q^2}{r^{2(N-2)}}
\right) dt^2+ \left(1-\frac{m}{r^{N-2}}+ \frac{q^2}{r^{2(N-2)}}
\right)^{-1} \,dr^2 \nonumber \\[4mm] &&
- \frac{2 \,a \sin^2 \theta}{r^{N-2}}\,\left(m
-\frac{q^2}{r^{N-2}}\right)\,dt \,d\phi + r^2 \left(d\theta^2
+\sin^2{\theta}\,d\phi^2 + \cos^2{\theta}  d\Omega_{N-3}^2\right)
 \label{desiredm}
\end{eqnarray}
where the parameter $ q $ is given by
\begin{equation}
q =\pm Q \left[\frac{8 \pi G}{(N-2)\,(N-1)\, A_{N-1}}\right]^{1/2}
\label{physcharge1}
\end{equation}
and $ A_{N-1} $ is the area of a unit $(N-1)$-sphere. The
associated electromagnetic field is described by the potential
one-form
\begin{equation}
A= -\frac{Q}{(N-2)\, r^{N-2}}\left(dt- a \sin^2\theta\,d\phi
\right)\,\,. \label{linpotform}
\end{equation}
The metric (\ref{desiredm}) generalizes the higher dimensional
Schwarzschild-Tangherlini \cite{tang} solution to include an
arbitrarily small angular momentum of the black holes. For
$\,N=4\,$ we have the metric in five dimensions. \cite{aliev2}.

It is clear that a rotating charged black hole must also have a
magnetic dipole moment. In our case,  it is determined from the
far distant behaviour of the  magnetic field in the metric
(\ref{desiredm}). We find that the associated magnetic
$\,(N-2)\,$-form field has the following orthonormal components
\cite{aliev1}
\begin{eqnarray}
B_{\hat r \,\hat\chi_{1} \hat\chi_2...\hat\chi_{N-3}} &=&
\frac{2\, Q a}{N-2}\,\,\frac{\cos\theta}{r^N}\,\,,~~~~~~~~~B_{\hat
\theta \,\hat\chi_1 \hat\chi_2...\hat\chi_{N-3}} = \frac{Q a
\sin\theta}{r^N}\,\,
\end{eqnarray}
which give the value of the  magnetic dipole moment $ \mu  = Q a
$. From the asymptotic behavior of the metric (\ref{desiredm}) we
find the specific angular momentum $ j= a m $  of the black hole.
The latter allows us to rewrite the magnetic dipole moment in
terms of total mass and total angular momentum of the black hole
as follows
\begin{eqnarray}
\mu&=&\frac{Q\,j}{m} = (N-1)\,\frac{Q\,J}{2 \,M}\,\,. \label{g1}
\end{eqnarray}
This expression  shows that a slowly rotating charged black hole
in $ N+1 $ dimensions must have the gyromagnetic ratio
\begin{equation}
g=N-1\,\,. \label{gyro}
\end{equation}
The value of the gyromagnetic ratio agrees with that obtained
earlier \cite{af} for the Myers-Perry black hole with a test
(small) electric charge in five dimensions. It also agrees with
the numerical analysis of paper \cite{kunz2}.

{\em Acknowledgements:} The author thanks the Scientific and
Technological Research Council of Turkey (T{\"U}B\.{I}TAK) for
partial support under the Research Project 105T437.

\vfill
\end{document}